\documentclass[twocolumn,showpacs,showkeys,superscriptaddress]{revtex4}
\usepackage{graphicx}
\usepackage{amsmath}
\usepackage{amsfonts}
\usepackage{amssymb}
\usepackage{mathrsfs}

\def\openone{\leavevmode\hbox{\small1\kern-3.8pt\normalsize1}}
\def\N{\leavevmode\hbox{ Z \kern-8 pt\normalsize{Z}}}
\def\openone{\leavevmode\hbox{\small1\kern-3.8pt\normalsize1}}
\def\openJ{\leavevmode\hbox{J \kern-9.5pt\normalsize J}}
\def\openS{\leavevmode\hbox{ S \kern-9.3pt\normalsize S}}
\newcommand{\bb}{\begin{equation}}
\newcommand{\ee}{\end{equation}}
\newcommand{\eqb}{\begin{eqnarray}}
\newcommand{\eqf}{\end{eqnarray}}

\usepackage{color}

\begin{document}

\title{Birefringent light propagation on anisotropic cosmological backgrounds}

\author{Felipe A. Asenjo}
\email{felipe.asenjo@uai.cl}
\affiliation{UAI Physics Center, Universidad Adolfo Ib\'a\~nez, Santiago, Chile.}
\affiliation{Facultad de Ingenier\'{\i}a y Ciencias,
Universidad Adolfo Ib\'a\~nez, Santiago, Chile.}
\author{Sergio A. Hojman}
\email{sergio.hojman@uai.cl}
\affiliation{UAI Physics Center, Universidad Adolfo Ib\'a\~nez, Santiago, Chile.}
\affiliation{Departamento de Ciencias, Facultad de Artes Liberales,
Universidad Adolfo Ib\'a\~nez, Santiago, Chile.}
\affiliation{Departamento de F\'{\i}sica, Facultad de Ciencias, Universidad de Chile,
Santiago, Chile.}
\affiliation{Centro de Recursos Educativos Avanzados,
CREA, Santiago, Chile.}

\begin{abstract}
Exact electromagnetic wave solutions to Maxwell equations on anisotropic Bianchi I cosmological spacetime backgrounds are studied.
The waves evolving on Bianchi I spacetimes exhibit birefringence (associated to linear polarization) and dispersion.
 The particular case of a vacuum--dominated anisotropic Universe, which reproduces a Friedmann-Robertson-Walker Universe (for late times) while for earlier times it matches a Kasner Universe, is studied. The electromagnetic waves do not, in general, follow null geodesics. This produces a modification of the cosmological redshift, which is now dependent on light polarization, its dispersion and its non-null geodesic behavior. New results presented here may help to tackle some issues related to the ``horizon'' problem.
\end{abstract}

\pacs{}

\maketitle

\section{Introduction}

Cosmology has been able to successfully pass the most demanding observational tests. However, there still are many deep open problems which remain to be solved \cite{pebles}. Several of them are related to early-Universe and to the Inflation era \cite{inflation}.  For such problems,  and for cosmology in general, it is customary to assume  that light propagates in null geodesics, implying a constant speed for the photon propagation.
In fact, it is very well-known that a plane electromagnetic (EM) wave propagating in any curved spacetime background  follows exactly a null geodesic in the high-frequency limit \cite{Whitt, misner}. This limit, also known as the eikonal or geometrical optic limit, is valid when the wavelength of the EM wave is much less than both the inverse of the square root of the typical component of the Riemann tensor measured by an observer in a local Lorentz frame and the typical length over which the wave features vary \cite{misner,mash1987}.
This analysis and description for a EM plane waves, widely used in problems concerning to light propagation in any curved space, seems to be satisfactory. Surprisingly, it is not. This work explore some unsatisfactory issues concerning to  the modification of the geodesics behavior of light propagating on anisotropic gravitational backgrounds.

Several problems related with the geodesic behavior of an EM wave arise when they are studied beyond the geometric optical limit. Different works along the past years have shown that the null geodesic behavior of light on curved spacetimes is not valid in general.
In 1960, DeWitt and Brehme \cite{dwb} proved that electromagnetic waves in a  curved spacetime background have a ``tail" radiation field inside the light cone, indicating a non-null path for the propagation of light. This has been also shown valid for the wave propagation of massless scalar fields in curved spacetimes \cite{dwb,Faraoni}. On the other hand, the geometrical optics approximation of Maxwell equations is sometimes superfluous while in other instances it may be plainly inadequate for waves whose frequencies do not satisfy the eikonal approximation assumptions, as it is the case for  plane waves that  are exact solutions of Maxwell equations for some specific metrics \cite{asenjohojmanLight}. These solutions do not impose any approximation on the scales of the wavelengths compared to the components of the Riemann tensor or the length over which the wave features vary. In particular, EM plane waves are exact solutions on a flat Friedmann-Robertson-Walker (FRW) background (see below).

Further, a non-null geodesic wave propagation
seems to be a sufficiently robust phenomenon to appear in the quantum realm for massless neutrinos   in curved spacetimes \cite{ohkuwa} and for Rarita-Schwinger waves in an external electromagnetic potential \cite{velo}.
 In addition to the above, it has been well-known for quite time ago that if vacuum quantum  effects (QED regime) are considered in the geometrical optics approximation of Maxwell equations in curved spacetime, the speed of  light is also modified, producing a superluminal propagation of an electromagnetic wave \cite{khri,hollow,Drummond,Klein,danielshore,danielshore2,danielshore3,danielshore4,Khriplovich},  showing birefringence \cite{heylH}, and problems associated to the non-conservation of the helicity of
a quantum electromagnetic fields \cite{agullo}. On the other hand, in the classical regime, it has  been shown that different models which describe the dynamics of relativistic spinning bodies give also rise to non--geodesic equations of motion \cite{mat,pap1,hojman1,gane1,gane2,armaza}.

That light, and other massless objects,  can move along paths which are not null geodesics (therefore at a speed which is not the speed of light)  seem to be shocking as they apparently violate the Equivalence Principle (EP) of General Relativity. However, it does not.
 It is, of course, correct to state that point particles (without any additional structures) follow geodesics when they move in the presence of gravitational forces only. Nevertheless, if one deals with extended (not point--like) objects, or with objects with additional structures such as spin, for instance, the geodesic path does not make sense. Several geodesic curves go across a sufficiently extended body, which is therefore subject to tidal forces.
 Waves are extended physical objects and thus they do not have to always propagate along null geodesics \cite{dwb}. In addition,  EM waves have spin, and wave polarization can also affect its propagation \cite{mashhoonPRD,evan,asenjohojmanLight}. It is important to emphasize that the EP cannot be used to fully understand the EM wave dynamics. If the EM wave scales are comparable to those of the gravitational fields, the wave does not travel along null geodesics. The EP is valid only in the high-frequency limit, where the EM wave is described as a classical massless particle that travels always along null geodesics \cite{ruizdodin}.

The above realization leads us to think that the EM wave propagation in curved spacetime beyond the geometrical optics limit is worth to be explored, understanding (in general) that any curved spacetime acts as an effective material medium where light propagates  \cite{asenjohojmanLight,plebanski,felice,mashhoon,mashhoon2,ruizdodin,mackay}.

It is the purpose of this work to study the behavior of light propagating on anisotropic cosmological backgrounds. The effects of spacetime anisotropy has been explored in the geometrical optics  limit before \cite{brans,sagnotti} finding that the anisotopy will induce rotations on the plane of polarization of the EM wave. It has been also studied the generation of linear polarization due to photon quantum effects in these metrics \cite{Moskaliuk}.
In contrast, here we study  the problem of the behavior of solutions of Maxwell equations, without any kind of quantum effects or different relativistic coupling formalisms.
 We show that solutions of the usual Maxwell equations on anisotropic cosmological backgrounds beyond the geometrical optics limit exhibit birefrigence and dispersion. In particular, EM wave solutions to Maxwell equations on gravitational anisotropic Kasner backgrounds \cite{kasner21} and on anisotropic vacuum--dominated universe \cite{ryan} metric exhibit the same features which appear on rotating backgrounds \cite{asenjohojmanLight}, i.e., birefringence, dispersion, and non-null geodesic propagation. All these effects are due to the coupling of curvature to the polarization of the EM waves.
We believe that the conclusions of this work allow for new approaches to tackle old problems.

\section{Maxwell equations on curved spacetime}
\label{sec2sec}

The dynamics of an EM field immersed in a gravitational background field is described by the Maxwell equations
\begin{equation}\label{MaxGeneralCurved}
\nabla_\alpha F^{\alpha\beta}=0\, ,\qquad \nabla_\alpha F^{*\alpha\beta}=0\,
\end{equation}
written in terms of  the antisymmetric electromagnetic field tensor $F^{\alpha\beta}$ (whose dual is $F^{*\alpha\beta}$), and where $\nabla_\alpha$ stands for the covariant derivative defined for a metric $g_{\mu\nu}$.
The previous equations can be studied in terms of the EM fields or by using the EM potentials. In the following we will describe both of them, as the former scheme is useful to identify the gravitational field as an effective medium, where the later one is more better suited to get solutions to the  wave equations.

In terms of electromagnetic fields, we can rewrite Eqs.~\eqref{MaxGeneralCurved} by defining
\begin{eqnarray}
E_i&=&F_{i0}\, ,\qquad D^i=\sqrt{-g} F^{0i}\, ,\nonumber\\
B^i&=&\varepsilon^{0ijk} F_{jk}\, , \quad\varepsilon^{0ijk}H_k=\sqrt{-g} F^{ij}\, ,
\end{eqnarray}
where $\varepsilon^{0ijk}$ is the Levi-Civita symbol and  $g$ the metric determinant (latin indices are used to denote space coordinates). Then, Eqs.~\eqref{MaxGeneralCurved} can be written simply as
\begin{eqnarray}\label{ecsMaxrgravMedium}
\partial_i D^i&=&0\, , \qquad \qquad\quad\partial_i B^i=0\, ,\nonumber\\
\partial_0 D^i&=&\varepsilon^{0ijk}\partial_j H_k\, , \quad \partial_0 B^i=-\varepsilon^{0ijk}\partial_j E_k\, ,
\end{eqnarray}
where $\partial_0$ stands for the time-derivative. These equations mimic the flat spacetime Maxwell equations for EM fields in a medium where \cite{plebanski}
\begin{eqnarray}
D^i&=&\epsilon^{ij}E_j-\varepsilon^{0ijk}\mu_j H_k\, \nonumber\\
B^i&=&\epsilon^{ij}H_j+\varepsilon^{0ijk}\mu_j E_k\, ,
\end{eqnarray}
with  ($g^{\mu\nu}$ is the inverse metric).
\begin{equation}\label{permitividadGrav}
\epsilon^{ij}=-\sqrt{-g}\frac{g^{ij}}{g_{00}}\, ,\qquad \mu_j=-\frac{g_{0j}}{g_{00}}\, .
\end{equation}
It is straightforward to realize that any gravitational field acts in the same way as a medium for EM wave propagation, with its corresponding permittivity and permeability. This can produce rotation of the polarization state of light \cite{goshsen}. In particular, for an anisotropic cosmology, $\mu_j=0$, but the  $\epsilon^{ij}$ eigenvalues (which may be functions of spacetime) are, in general, different in the three spatial directions (implying birefringence).

The study of EM waves using Eqs.~\eqref{ecsMaxrgravMedium} requires to increase the derivative order of the Maxwell equations. To avoid that, it is more useful to express the electromagnetic field in terms of the four-vector potentials $A_\alpha$, such that $F^{\alpha\beta}=g^{\alpha \mu} g^{\beta \nu}F_{\mu \nu}$, and
\begin{equation}
F_{\mu\nu}=\nabla_\mu A_\nu-\nabla_\nu A_\mu\equiv\partial_\mu A_\nu-\partial_\nu A_\mu\, ,
\end{equation}
 where $\partial_\mu$ is a partial derivative. Hence, from \eqref{MaxGeneralCurved}, we see that the equations  $\nabla_\alpha F^{*\alpha\beta}=0$ are identically satisfied, whereas the only equations which remain to be solved are
\begin{equation}\label{eMax1}
\frac{1}{\sqrt{-g}}\partial_\alpha\left[\sqrt{-g}g^{\alpha\mu}g^{\beta\nu} (\partial_\mu A_\nu-\partial_\nu A_\mu)\right]=0\, .
\end{equation}
Eqs.~\eqref{eMax1} are the ones to be solved along this work, as they show below in a clear fashion the birefringence produced by anisotropic spacetimes.

In general, there are several exact solutions to Maxwell equations \cite{asenjohojmanLight,plebanski,teu,chandra,felice,mashhoon,mashhoon2,cohen,tsagas}. One of the most interesting solutions of Eqs.~\eqref{eMax1} are EM plane waves  \cite{misner,born} that can be represented by
\begin{equation}\label{formaApot}
A_\mu=\xi_\mu e^{i{S}}\, ,
\end{equation}
  where $\xi_\mu$  and ${S}$ represent the amplitude and the  phase of the wave. We can choose both quantities to be real functions of space and time. The wavevector of the wave is defined by
\begin{equation}
K_\mu=\nabla_\mu { S}=\partial_\mu { S}\, ,
\end{equation}
where $K_0$ is the frequency of the wave and $K_i$ are the components of the (three dimensional) wavevector.

The nature of the propagation of an EM wave lies in the values of the scalar $K_\mu K^\mu$. This scalar is known to yield the dispersion relation of the wave, and every EM wave  that follows null geodesics must satisfy  $K_\mu K^\mu\equiv 0$.
In vacuum flat-spacetime, Eqs.~\eqref{eMax1} are solved by constant amplitude and $K_\mu K^\mu=0$, implying that all EM plane waves in flat spacetime evolve along null geodesics, travelling at the speed of light. In this case, we can see from \eqref{permitividadGrav} that $\epsilon^{ij}=\eta^{ij}$ and $\mu_j=0$ (with $\eta^{ij}$ the flat spacetime metric) and thus, there is no effective medium.
On the other hand, on curved spacetimes, it is customary to solve the general Maxwell equations \eqref{eMax1}  using the geometrical optics approximation \cite{plebanski,felice,misner,born,tolman}, through a WKB like scheme. Under this approximation, the scales of variations of the amplitude and of the gravitational field are neglected, and Eqs.~\eqref{eMax1} will always  describe transversal EM plane waves following null geodesics. This approximated solution is usually invoked to claim that all EM waves travel in null geodesics in any arbitrary gravitational background. As we prove here, that fact does not hold for general spacetimes.

In general, withouth any approximation, Eqs.~\eqref{eMax1} can be casted in a set of covariant equations that explicitly describe the non-null behavior of an EM plane wave. If the wave is transversal, the condition
\begin{equation}\label{transversalcondition}
K_\mu \xi^\mu=0\, ,
\end{equation}
must hold. Moreover, by choosing the Lorentz gauge  $\nabla_\mu A^\mu=0$, then we also conclude that
\begin{equation}
\nabla_\mu \xi^\mu=0\, ,
\end{equation}
for the particular solution of these EM plane waves with real amplitude and phase.

In this way, using \eqref{formaApot} for the electromagnetic vector potential, Eqs.~\eqref{eMax1} becomes
\begin{equation}\label{SeparadaMax1}
\left(K^\mu K_\mu\right) \xi^\beta=\frac{1}{\sqrt{-g}}\partial_\alpha\left[\sqrt{-g}g^{\alpha\mu}g^{\beta\nu}\left(\partial_\mu \xi_\nu-\partial_\nu\xi_\mu\right)\right]\, ,
\end{equation}
and
\begin{eqnarray}\label{SeparadaMax2}
0&=&\partial_\alpha\left[\sqrt{-g}g^{\alpha\mu}g^{\beta\nu}\left(K_\mu \xi_\nu-K_\nu\xi_\mu\right)\right]\nonumber\\
&&+\sqrt{-g}K^\mu g^{\beta\nu}\left(\partial_\mu \xi_\nu-\partial_\nu\xi_\mu\right)\, ,
\end{eqnarray}
where the transversality condition \eqref{transversalcondition} has been used.

Eq.~\eqref{SeparadaMax1} gives origin to the dispersion relation
\begin{equation}\label{SeparadaMax3}
K^\mu K_\mu =\chi\, ,
\end{equation}
where (with $\xi\equiv\sqrt{\xi_\mu\xi^\mu}$)
\begin{equation}\label{SeparadaMaxGG3}
\chi =\frac{\xi_\beta}{\sqrt{-g}\, \xi^2}\partial_\alpha\left[\sqrt{-g}g^{\alpha\mu}g^{\beta\nu}\left(\partial_\mu \xi_\nu-\partial_\nu\xi_\mu\right)\right]\, ,
\end{equation}
does not vanish, in general.
It is straightforward to check that in the flat spacetime limit, the constant amplitude ansatz produces null geodesics $K^\mu K_\mu =0$. Also, in the geometrical optics limit, the right-hand side is neglected, again modeling light propagation in null geodesics. However, beyond that limit, is very unlikely that the right-hand side vanishes, as the constant amplitude assumption is incompatible with Eq.~\eqref{SeparadaMax2}, in general. This means that the wave does not follow null geodesics, as from Eq.~\eqref{SeparadaMax3} we obtain that
\begin{equation}\label{SeparadaMaxGG}
K^\nu\nabla_\nu K_\mu=\frac{1}{2}\partial_\mu \chi\, ,
\end{equation}
where we have used that $\nabla_\mu K_\nu=\nabla_\nu K_\mu$. Eq.~\eqref{SeparadaMaxGG} is a consequence of the spreading (not point-like character) of the wave.

On the other hand, Eq.~\eqref{SeparadaMax2} can be exactly rewritten as
\begin{eqnarray}\label{SeparadaMax4}
2K^\mu\nabla_\mu \xi^\beta+\left(\nabla_\mu K^\mu\right) \xi^\beta=0\, ,
\end{eqnarray}
for the Lorentz gauge. This equation can also be derived in the geometrical optics limit \cite{misner}. However,  here we show that this equation is exact for transverse waves on any curved spacetime.
A remarkable and important feature can also be obtained from Eq.~\eqref{SeparadaMax4}. Contracting it with $\xi_\beta$, it implies the conservation of photon number in general (without using the WKB approximation)
\begin{eqnarray}\label{SeparadaMax5}
\nabla_\mu\left(\xi^2 K^\mu\right)= 0\, .
\end{eqnarray}
This result holds for Maxwell equation on any curved spacetime beyond any approximation on the scales of the wavelength, contrary to what is usually invoked \cite{misner}.

Finally, by using Eqs.~\eqref{SeparadaMax4} and \eqref{SeparadaMax5}, we can show that the polarization vector $f^\mu=\xi^\mu/\xi$ follows
\begin{equation}\label{SeparadaMax6}
K^\mu\nabla_\mu f^\beta=0\, ,
\end{equation}
in the Lorentz gauge. Using the above equation and Eq.~\eqref{SeparadaMaxGG}  we can also find
\begin{equation}\label{SeparadaMax7}
K^\mu\nabla_\mu\left(K_\beta f^\beta\right)=\frac{1}{2}f^\beta \partial_\beta\chi\, .
\end{equation}
If the right-hand side vanishes, this equation predicts that the polarization vector of the wave is parallel-propagated along the rays \cite{misner}. Otherwise, it is not. For example, for the propagation in an anisotropic cosmological background, the EM waves fulfill $f^\beta \partial_\beta\chi\equiv 0$ (see below), and the polarization of the waves is parallel-propagated along the rays for those solutions.

Thereby, Eqs.~\eqref{SeparadaMax1} and \eqref{SeparadaMax4} allow us to calculate the plane wave solution of Maxwell equation in general. Equations ~\eqref{SeparadaMax5} and \eqref{SeparadaMax6} assure us that they correspond to physical waves propagating on a general background spacetime.

%%%%%%%%%%%%%%%%%%%%

\section{Propagation of light in anisotropic universes}
\label{Propagationoflightinanisotropicsec}

Consider the  Bianchi I spacetime interval  in cartesian coordinates  \cite{ryan}
\begin{equation}\label{animetric}
ds^2=-dt^2+a(t)^2 dx^2+b(t)^2 dy^2+c(t)^2 dz^2\, ,
\end{equation}
 where $a$, $b$ and $c$, are three arbitrary functions of time. Maxwell equations have now time-dependent permittivity and permeability coefficients $\epsilon^{11}={bc}/{a}$, $\epsilon^{22}={ac}/{b}$,  $\epsilon^{33}={ab}/{c}$, and $\mu_i=0$.
It is straightforward to realize that EM wave propagation depends on the direction, as $\epsilon^{11}\neq \epsilon^{22}\neq\epsilon^{33}\neq\epsilon^{11}$ in general. Therefore, the anisotropic cosmological background behaves as an  effective anisotropic birefringent medium for EM plane waves.
For every direction of propagation, there are two polarizations that propagate differently.
For simplicity, from now on, we consider EM vector potentials which depend on time and one spatial coordinate only.

Let us start with an EM wave propagating in the $x$-direction. We  consider an EM potential $A_\mu(t,x)$ with neither $y$ nor $z$ dependence.   That spatial coordinate indicates the direction of propagation of the EM wave.
Maxwell Eqs.~\eqref{eMax1} become
\begin{eqnarray}
&&-\partial_0\left[abc\,  g^{\mu\nu}\left(\partial_0 A_\nu-\partial_\nu A_0\right)\right]\nonumber\\
&&\qquad\qquad+\frac{bc}{a}\partial_x\left[g^{\mu\nu}\left(\partial_x A_\nu-\partial_\nu A_x\right)\right]=0\, .
\end{eqnarray}
We write Maxwell equations for $\mu = 0, 1, 2, 3 (= \nu)$. For  $\mu = \nu = 0$ and $\mu=\nu=1$, the equations are
\begin{eqnarray}
\partial_x \left[\frac{b c}{a}( \partial_x A_0 - \partial_0 A_x) \right]&=&0\, ,\label{xA0}\\
\partial_0 \left[\frac{b c}{a}( \partial_x A_0 - \partial_0 A_x) \right]&=&0\, ,\label{xAx}
\end{eqnarray}
implying that the longitudinal component of the electric field
\begin{equation}
\partial_x A_0 - \partial_0 A_x  = \frac{\zeta\, a}{b c}\, ,
\end{equation}
is given in terms of an arbitrary constant $\zeta$. Consequently, the longitudinal electric field  may vanish by choosing $\zeta=0$.  The simplest solution for the longitudinal fields is to choose $A_x=0=A_0$. This choice assures us that the wave to be described fulfills the transversality condition and that the wave moves always parallel-propagated along the rays. Furthermore, these choices implies that the electromagnetic potential vector satisfies identically the Lorenz gauge.

Moreover, the equations for $\mu = \nu = 2$, and $\mu = \nu = 3$, which determine $A_y (t,x)$ and $A_z (t,x)$, are uncoupled to the previous ones. They respectively  are
\begin{eqnarray}
-\frac{ab}{c}\partial_0\left(\frac{ac}{b}\partial_0 A_y\right)+\partial_x^2A_y&=&0\, ,\label{xAy}\\
-\frac{ac}{b}\partial_0\left(\frac{ab}{c}\partial_0 A_z\right)+\partial_x^2A_z&=&0\, .\label{xAz}
\end{eqnarray}
Eqs.~\eqref{xAy} and \eqref{xAz} do not coincide, in general. By the previous choices, they represent EM plane waves propagating in the $x$-direction in our coordinates. From these equations, it is direct to notice that the wave propagations for polarizations $A_y (t,x)$ and $A_z (t,x)$ are refracted differently, and thus, birefringence is exhibited in a general anisotropic cosmology.

Similar equations for the transverse components of Maxwell equations \eqref{eMax1}
can be straightforwardly found for the wave propagation along the $y$-direction (with either $x$ or $z$-polarization) or  along the $z$-direction (with $x$ or $y$-polarization), when the longitudinal vector potentials are chosen to vanish. We
can write a general expression for the six plane waves propagating in any direction for every polarization as
\begin{equation}
-\frac{a_i a_k}{a_j}\partial_0\left(\frac{a_j a_k}{a_i}\partial_0 A_{i}(t,x^k)\right)+\frac{\partial^2A_{i}(t,x^k)}{\partial {x^k}^2}=0\, ,\label{yAxG}
\end{equation}
with $i,j,k=1,2,3$ and $i\neq j\neq k\neq i$, $a_1=a$, $a_2=b$, and $a_3=c$,  $x^1=x$, $x^2=y$, and $x^3=z$.

The wave equations \eqref{yAxG} are not the same in general.
 Different polarizations  propagating in different directions  behave differently thus producing birefrigence.
 This strongly suggests that a fully or partially anisotropic spacetime induces birefrigence for the different polarizations of the wave and for different directions where it propagates.

The solutions to Eqs.~\eqref{yAxG} are EM waves such their plane of oscillation is always perpendicular to the direction of propagation, as in the case of vacuum flat spacetime.
Solutions to Eqs.~\eqref{yAxG} for particular anisotropic spacetimes are discussed in Sec.~\ref{soluAnis}. Nonetheless, before proceeding with the analysis of the  solutions, several features of thes equations deserve to be discussed in detail.

%%%%%%%%%%%%%%%%%%%%%%%%%%
\subsection{Isotropic limit}

When $a=b=c$, the spacetime becomes isotropic. In this case,
all equations  describing transverse modes of  Eqs.~\eqref{yAxG} coincide in any generic direction $\vartheta$
 \begin{equation}
-a\partial_0\left({a}\partial_0 A\right)+\partial_\vartheta^2A=0\, .
\end{equation}
 All polarizations propagate in the same fashion and no birefringence is displayed. Every solution has the form of plane waves $\exp(iS)$, with constant amplitude and phase
\begin{equation}
S= k (\vartheta\pm\tau_{\text{FRW}})\, ,
\end{equation}
where $k$ is a constant and
\begin{equation}
\tau_{\text{FRW}}=\int \frac{dt}{a}\, ,
\end{equation}
is the FRW time \cite{asenjohojmanLight}.  From the phase we can calculate the frequency and wavevector \cite{asenjohojmanLight}
\begin{equation}
K_0=\partial_0 S=\frac{k}{a}\, ,\qquad K_\vartheta=\partial_\vartheta S=k\, .
\end{equation}
These two quantities allow us to obtain that,  in a flat FRW universe, light  propagates along null geodesics defined by $K_\mu K^\mu\equiv 0$.

Notice that this is an exact solution of Maxwell equations for flat FRW spacetime. In fact, plane EM waves are an exact solution of Maxwell equations for conformally flat spacetimes \cite{noonan}.
 No assumptions on the scale of the wave are needed to obtain this solution.

%%%%%%%%
\subsection{Impossibility to get exact solutions which follow null geodesics for anisotropic spacetimes}

In general, an exact EM plane wave solution to  Maxwell equations on an anisotropic background cannot follow null geodesics (this  can occur only in the high-frequency limit, see next section). This can be directly proved by using  the wave equations \eqref{yAxG}. These equations share the same structure for  EM wave propagation in a generic direction $\vartheta$
\begin{equation}
0=-{p}\, h\partial_0\left(\frac{p}{h}\partial_0 A\right)+\partial_\vartheta^2A\, ,
\end{equation}
where $p$ and $h$ are functions of time.

Let us assume a plane wave solution $A(t,\alpha)=\xi(t)\exp[iS(t,\vartheta)]$, with $S(t,\vartheta)=s(t)+k\vartheta$, where $k$ is a constant, and $\xi$ and $s$ are real functions.
In order for this expression to be a plane wave solution, the functions must satisfy
\begin{equation}\label{nullcondincon}
p^2 (\partial_0 s)^2-k^2=\frac{p h}{\xi}\partial_0\left(\frac{p\partial_0\xi}{h}\right)\, ,\quad
\partial_0\left(\frac{p}{h}\xi^2\partial_0 s\right)=0\, ,
\end{equation}
where the first equation is the dispersion relation of the wave \eqref{SeparadaMax3}, and the second one is the conservation of photons \eqref{SeparadaMax5}. The null geodesic condition for the wave is
\begin{equation}
K_\mu K^\mu=-(\partial_0 s)^2+k^2/p^2=0\, ,
\end{equation}
 where $K_0=\partial_0 S=\partial_0 s$, $K_\alpha=k$, and $g^{\alpha\alpha}=p^{-2}$.  Therefore, if we demand that the plane wave travels in null geodesics, then $\partial_0 s={k}/{p}$, and the amplitude $\xi$ must satisfy both Eqs.~\eqref{nullcondincon}, that forces the condition
\begin{equation}\label{condinconstetten}
\frac{p}{h^{3/2}}\partial_0 h=\text{constant}\, .
\end{equation}
Obviously, this condition does not hold for general $p$ and $h$, and thus, null geodesic behavior is not a general solution of wave equations \eqref{yAxG}. A trivial solution of \eqref{condinconstetten}
occurs when $h$ equal to a constant that correspond to the isotropic flat FRW case.

%%%%%%%%%%%%%%%%%%%%%%%%
\subsection{High-frequency limit}
\label{sechighFF}

As was mentioned before, the geometrical optics (or eikonal  limit) \cite{misner} implies that light propagates always along null geodesics, travelling at the speed of light. The geometrical optics regime corresponds to the high-frequency limit of the wave equations \eqref{yAxG}. This limit occurs when the variation of the amplitude of the wave are  small compared to the variation scales of the phase. For example, the plane wave solutions of Eqs.~\eqref{xAy} and \eqref{xAz}, in the  $x$--propagation, are simply
\begin{eqnarray}\label{highfreqSol}
A_y(t,x)&\propto& \sqrt{\frac{b}{c}}\exp\left(i S_x\right)\, ,\nonumber\\
A_z(t,x)&\propto&\sqrt{\frac{c}{b}}\exp(iS_x)\, ,
\end{eqnarray}
where $S_x=k(\tau_x\pm x)$, and  $\tau_x=\int dt/a$ is a ``FRW time variable'' along the $x$--direction. The high-frequency limit is achieved under the assumption that amplitude variations are negligible compared to the wave frequency,  $(ab/c)\partial_t\left[(ac/b)\partial_t \xi_{y,z} \right]/\xi_{y,z}\ll  ( \partial_t  S_x)^2$.
 Defining the frequency $K_0\equiv\partial_t S_x =k/a$ and the  wavevector $K_x \equiv \partial_x S_x=\pm k$, then along the $x$--direction $K_\mu K^\mu=0$ is fulfilled.
Even though the wave propagates along null geodesics, the amplitude of each polarization is different and time-dependent. Thereby the electric and magnetic fields associated to the above potentials have different time-dependent amplitudes \cite{brans,sagnotti}, and the polarization of the EM wave can  rotate.

Similar solutions in the high--frequency limit can be found for other directions of propagation of the EM waves, with similar conclusions.

In the high-frequency limit, the EM wave travels along null geodesics in any direction, but with different amplitudes and time variables for different polarizations. The ``speed of light'' propagation result for EM waves is achieved at the price of defining  different time variables in different space directions.

The importance of results presented in this section, for instance solutions \eqref{highfreqSol}, is that they can be obtained as limits of the general solutions of the following section, as we will show.

%%%%%%%%%%%%%%%%%%%%%%%%%%%%%

\section{A simple theoretical example}
\label{modelteoricsimple}

Let us consider a general simple case to exemplify the differences that emerges in anisotropic universes compared to isotropic ones. Assume that the metric is in the form $a(t)=c(t)$, and $b(t)=a(t)[1+\epsilon(t)]$, where $\epsilon\ll1$. Thus the spacetime has a small anisotropy in one direction.

We can study the propagation of light in any direction. We can start with the propagation on the $y$-direction such that the wave amplitude is polarized in the $x$-direction $\xi_x(t)$ and depends on time only. The non-zero four-wavevector components are $K_0(t)$ and $K_y=k$, where $k$ is a constant. These conditions preserve transversality.
The wave equation to describe this propagation is contained in Eqs.~\eqref{yAxG}. However, from the conservation of photons \eqref{SeparadaMax5} is direct to obtain that
\begin{equation}
\xi_x=\frac{\xi_{0x}}{\sqrt{b K_0}}\, ,
\end{equation}
where $\xi_{0x}$ is a constant.
It is not difficult to prove that this is also a solution of Eq.~\eqref{SeparadaMax4}. Using this amplitude, we can now obtain the dispersion relation from \eqref{SeparadaMax1} as
\begin{eqnarray}
\frac{-K_0^2+k^2/b^2}{\sqrt{b K_0}}=-\frac{1}{b}\partial_0\left[b\partial_0\left(\frac{1}{\sqrt{b K_0}}\right)\right]\, .
\end{eqnarray}
From here is clear that the simplest solution is $b K_0=k$, which makes both sides of the equation to vanish identically. This solution implies that the propagation in this direction follows null geodesics $K_\mu K^\mu\equiv -K_0^2+k^2/b^2=0$, with constant amplitude waves. The polarization in the $z$-direction has the same dynamics for its propagation.

A different story is what happens for waves propagating in the $x$-direction, with non-zero four-wavevector components $K_0(t)$ and $K_x=k$ (with constant $k$). First assume a $z$-polarized wave with amplitude $\xi_z$, such that from Eq.~\eqref{SeparadaMax5} we find
\begin{equation}\label{solTeoejem0a}
\xi_z=\frac{\xi_{0z}}{\sqrt{b K_0}}\, ,
\end{equation}
where $\xi_{0z}$ is again a constant. This solution also solves Eq.~\eqref{SeparadaMax4}. The amplitude now depends on time.
With this amplitude, from the dispersion relation \eqref{SeparadaMax1} we get
\begin{eqnarray}\label{solTeoejem0}
-K_0^2+\frac{k^2}{a^2}=-\frac{\sqrt{bK_0}}{b}\partial_0\left[b\partial_0\left(\frac{1}{\sqrt{b K_0}}\right)\right]\, .
\end{eqnarray}
If these waves move along null geodesics, then $K_0=k/a$ should be a solution of the previous equation. However, it is not, and the waves do not move in such way. An approximate solution can be found if we assume that
\begin{equation}\label{solTeoejem1}
K_0=\frac{k}{a}(1+\eta_z)\, ,
\end{equation}
where $\eta_z\ll 1$ is the  correction due to spacetime anisotropy. Using \eqref{solTeoejem1} in \eqref{solTeoejem0}, we find that
\begin{eqnarray}\label{solTeoejem2}
\eta_z(\tau,k)&=&\frac{1}{2k}\cos(2k\tau)\int_0^\tau \frac{\partial^2\epsilon(\tau')}{\partial{\tau'}^2}\sin(2k \tau') d\tau'\nonumber\\
&&-\frac{1}{2k}\sin(2k\tau)\int_0^\tau\frac{\partial^2\epsilon(\tau')}{\partial{\tau'}^2}\cos(2k \tau') d\tau'\, ,
\end{eqnarray}
with $\tau=\int dt/a$. Importantly, as $\epsilon\neq 0$ represents the anisotropy, then $\eta\neq 0$ in general. Thus, the wave follows a non-null path described by
\begin{equation}
K_\mu K^\mu\approx-\frac{2k^2}{a^2}\eta_z\, ,
\end{equation}
which could be either timelike or spacelike. Besides, the wave now is dispersive as its dispersion relation depends on $k$.
 Notice that if the Universe isotropizes, then the frequency \eqref{solTeoejem1}  approaches its value in the high-frequency limit (Sec.~\ref{sechighFF}) when $\epsilon\rightarrow 0$, as for the oscillating
function \eqref{solTeoejem2} averages to zero. Besides, the amplitude \eqref{solTeoejem0a}  behaves as $\sqrt{c/b}(1-\eta_z/2)$,  that approaches the wave amplitude \eqref{highfreqSol} in the high--frequency limit.

Now, let us study the $y$-polarized wave with amplitude $\xi_y$. By using Eq.~\eqref{SeparadaMax5} we obtain that
\begin{equation}\label{aplxiysec4a}
\xi_y=\xi_{0y}\sqrt{\frac{b}{a^2 K_0}}\, ,
\end{equation}
where $\xi_{0z}$ is again a constant.  The $y$-polarized wave has a different amplitude than the $z$-polarized one. Both of them depend on time, and therefore the plane of polarization rotates.
With the above amplitude, from \eqref{SeparadaMax1} we get
\begin{eqnarray}\label{solTeoejem4}
-K_0^2+\frac{k^2}{a^2}=-\frac{\sqrt{bK_0}}{a}\partial_0\left[\frac{a^2}{b}\partial_0\left(\sqrt{\frac{b}{a^2 K_0}}\right)\right]\, .
\end{eqnarray}
The null geodesic path $K_0=k/a$ is not a solution again. Assuming the approximated solution
\begin{equation}\label{solTeoejem5}
K_0=\frac{k}{a}(1+\eta_y)\, ,
\end{equation}
with $\eta_y\ll 1$, then using \eqref{solTeoejem4} we find
\begin{eqnarray}\label{solTeoejem6}
\eta_y(\tau,k)&=&-\frac{1}{2k}\cos(2k\tau)\int_0^\tau \frac{\partial^2\epsilon(\tau')}{\partial{\tau'}^2}\sin(2k \tau') d\tau'\nonumber\\
&&+\frac{1}{2k}\sin(2k\tau)\int_0^\tau\frac{\partial^2\epsilon(\tau')}{\partial{\tau'}^2}\cos(2k \tau') d\tau'\, ,
\end{eqnarray}
and the wave does not follow null geodesics as
\begin{equation}
K_\mu K^\mu=-\frac{2k^2}{a^2}\eta_y\ \neq \ 0, .
\end{equation}
Anew, if the Universe isotropizes, the frequency \eqref{solTeoejem5} approaches to its value in the high-frequency limit when $\epsilon\rightarrow 0$ (Sec.~\ref{sechighFF}). Also, the amplitude \eqref{aplxiysec4a} is  proportional to $\sqrt{b/c}(1-\eta_y/2)$, approaching to the wave amplitude \eqref{highfreqSol} in the high--frequency (isotropic) limit.

From \eqref{solTeoejem2} and \eqref{solTeoejem6}, we can also see that these two linear polarizations fulfill $\eta_z=-\eta_y$, and therefore
\begin{equation}
\left.K_\mu K^\mu\right|_{\text{$z$-polarized}}=-\left.K_\mu K^\mu\right|_{\text{$y$-polarized}}\, .
\end{equation}
Therefore, for the case of a small anisotropy, while one polarization moves along timelike curves, the other one will follow spacelike ones.

%%%%%%%%%%%%%%%%%%%%%%%

\section{Anisotropic vacuum-dominated universe}
\label{soluAnis}

As we previously discussed, Maxwell equations  \eqref{yAxG} can be studied and solved for arbitrary frequencies (not only in the geometrical optics limit).
In principle, these solutions of  Eqs.~\eqref{yAxG} show that  plane waves  do not propagate along null geodesics.
For the sake of clarity, let us explore the consequences of  Eqs.~\eqref{yAxG} in a specific anisotropic  cosmological model for an universe filled with a  perfect fluid whose energy density corresponds to a positive cosmological constant $\Lambda$. In this solution, the spacetime metric \eqref{animetric} has components \cite{Gron}.
\begin{eqnarray}\label{metricKacomp}
a(t)&=&[\beta(t)]^{1/3}[\Theta(t)]^{\frac{2}{3}\cos(\phi+\pi/3)}\, ,\nonumber\\
b(t)&=&[\beta(t)]^{1/3}[\Theta(t)]^{\frac{2}{3}\cos(\phi-\pi/3)}\, ,\nonumber\\
c(t)&=&[\beta(t)]^{1/3}[\Theta(t)]^{-\frac{2}{3}\cos(\phi)}\, ,
\end{eqnarray}
where $\beta(t)=\sinh(\sqrt{3\Lambda} t)$, $\Theta(t)=\tanh(\sqrt{3\Lambda} t/2)$, and $\phi$ is a constant. This model  asymptotically gives rise  to isotropization of the Universe, reproducing for large times a FRW cosmology with constant Hubble parameter (meaning a that the Universe is more anisotropic at earlier times).

The model \eqref{metricKacomp} can be fully anisotropic for arbitrary values of $\phi$. To show, in the simplest way, how anisotropy modifies light propagation, we choose $\phi=\pi/3$. In this case, the model is simplified as $a=c\neq b$, and the metric is now anisotropic in the $y$-direction. Therefore we expect that the EM waves propagating in $x$ and $z$-directions behave in the same form, but differently from those which propagate in the $y$-direction.
This is similar to the case that we studied in the previous section, as
\begin{equation}\label{comparisonab}
\frac{b}{a}=\Theta\neq 1\, .
\end{equation}

\subsection{Propagation in the $y$-direction}

The two polarizations $A_x(t,y)$ and $A_z(t,y)$ [collectivelly denoted by $A_\rho(t,y)$] of an EM wave  propagating along the $y$-direction, described by Eqs.~\eqref{yAxG}, fulfill now the same equation
\begin{equation}\label{vaccumAy2}
-\beta^{1/3}\Theta^{\frac{2}{3}}\partial_0\left(\beta^{1/3}\Theta^{\frac{2}{3}}\partial_0 A_\rho\right)+\partial_y^2 A_\rho=0\, .
\end{equation}
The solution of the above equation is a plane wave with the form $A_\rho(t,y)=\exp[ik(y\pm \tau_y)]$, with  a constant $k$, and
\begin{equation}\label{timenulmahemaia}
\tau_y=\int \beta^{-1/3}\Theta^{-\frac{2}{3}} dt\, .
\end{equation}
Consequently, these waves move along null geodesics with $K_\mu K^\mu=-K_0^2+g^{yy}K_y^2=0$, and
\begin{equation}
K_0=k\partial_0\tau_y=k \beta^{-1/3}\Theta^{-\frac{2}{3}}\, ,\quad K_y=k\, .
\end{equation}

The null geodesic behavior of this mode can be mathematically understood considering the time definition \eqref{timenulmahemaia}. This time allow us to put the Eq.~\eqref{vaccumAy2} in a flat spacetime-like form. Whenever this is not possible, the propagation of light does not take place along null geodesics.

\subsection{Propagation in the $x$-direction}

Waves propagating in the $x$-direction can have two different linear polarizations. We study each case in the limit of very early and large times. The wave propagation in the $z$-direction behaves analogously.

\subsubsection{$z$-polarized EM wave}

A different behavior is exhibited by a $z$-polarized EM wave which propagates in the $x$-direction, described by Eq.~\eqref{xAz}. In this case, the wave equation is
\begin{equation}\label{vaccumAy}
-\beta^{1/3}\Theta^{-\frac{4}{3}}\partial_0\left(\beta^{1/3}\Theta^{\frac{2}{3}}\partial_0 A_z\right)+\partial_x^2 A_z=0\, .
\end{equation}
For this wave, the null geodesics propagation is not an exact solution.
The most  interesting limiting cases of the previous equation are late times (where this cosmology  approaches to an isotropic FRW model with exponential expansion) and very early times (where the model approaches  an anisotropic Kasner vacuum Universe).

First, for late times $t\rightarrow \infty$, then $\beta\sim \exp(\sqrt{3\Lambda} t)/2$, and $\Theta\sim 1-2\exp(-\sqrt{3\Lambda} t)$.  In this case $b\approx a(1+\epsilon)$, where from Eq.~\eqref{comparisonab} we obtain
\begin{equation}\label{solepsilongggg1}
\epsilon\approx-2e^{-\sqrt{3\Lambda}t}\, .
\end{equation}
Thereby, this case can be treated as the model of Sec.~\ref{modelteoricsimple}, where $A_z=\xi_z \exp(iS)$. The new time for this case is
\begin{equation}
\tau=\int_t^\infty \frac{dt}{a}=2^{1/3}\sqrt{\frac{3}{\Lambda}}e^{-\sqrt{3\Lambda}t/3}\left(1-\frac{1}{6}e^{-\sqrt{3\Lambda}t}\right)\, ,
\end{equation}
 that approaches  $\tau_{\text{FRW}}$ as $t\rightarrow\infty$. In this limit, we can write \eqref{solepsilongggg1} as
\begin{equation}\label{solepsilongggg2}
\epsilon\approx-\left(\frac{\Lambda}{3}\right)^{3/2}\tau^3\, .
\end{equation}
This allow us to find the correction to the frequency due to curvature \eqref{solTeoejem2}
\begin{equation}
\eta_z\approx \left(\frac{\Lambda}{3}\right)^{3/2}\tau^3- \left(\frac{\Lambda}{3}\right)^{3/2}\frac{k^2\tau^5}{5},
\end{equation}
 which, in turn, allow us to find the amplitude \eqref{solTeoejem0a}, and the frequency \eqref{solTeoejem1}
\begin{eqnarray}
\xi_z&\approx&\frac{\xi_{0z}}{\sqrt{k}}\left[1+ \left(\frac{\Lambda}{3}\right)^{3/2}\frac{k^2\tau^5}{10}\right]\, ,\nonumber\\
K_0&=&\partial_0 S\approx \frac{k}{a}\left[1+\left(\frac{\Lambda}{3}\right)^{3/2}\tau^3\right]\, ,
\end{eqnarray}
where the wavevector is $K_x=\partial_x S=k$.
 With these results, we can obtain the  behavior of this wave
\begin{equation}\label{disp1late}
K_\mu K^\mu\approx-\frac{2}{a^2}\left(\frac{\Lambda}{3}\right)^{3/2}k^2\tau^3<0\, .
\end{equation}
This solution has a timelike nature, implying that a $z$-polarized the EM plane wave propagating in the $x$-direction in this anisotropic cosmology behaves as massive particles travelling at speed lower than the speed of light.  Notice that the wave  disperses as its frequency depends on the wavevector.
As $t\rightarrow \infty$ ($\tau,\eta_z\rightarrow 0$), corresponding to the isotropic FRW limit, the EM plane wave approaches a null geodesics behavior $K_\mu K^\mu=0$.

Now one can wonder what happens in the very early time case, when $t\rightarrow 0$. In this case,
$\beta(t)\sim \sqrt{3\Lambda}t$, and $\Theta(t)\sim \sqrt{3\Lambda}t/2$. This metric corresponds to the Kasner vacuum cosmology which describes an anisotropic Universe with vanishing energy-momentum matter tensor \cite{kasner21}.
The $z$-polarized EM wave propagating in the $x$-direction, depicted by  Eq.~\eqref{vaccumAy}, now satisfies
\begin{equation}\label{vaccumAyearly}
-\frac{2^{2/3}}{t}\partial_0\left(t\, \partial_0 A_z\right)+\partial_x^2A_z=0\, ,
\end{equation}
which has the exact solution
\begin{equation}\label{solHankel0small}
A_z(t,x)=A_{z0}\, H_0^{(1)}\left(2^{-1/3} k t\right) e^{\pm i k x}\, ,
\end{equation}
where $A_{z0}$ is an arbitrary constant, $H_0^{(1)}=J_0+iY_0$ is the Hankel function of zeroth order, and $J_0$ and $Y_0$ are the zeroth order Bessel functions of first and second kind respectively. Similar results have been found in Ref.~\cite{sagnotti}.

For very small times $k\, t\ll 1$, solution \eqref{solHankel0small} has the approximated form of a plane wave $A_z(t,x)=A_{z0}\xi(t) e^{i S(t,x)}$, where the amplitude and the phase are respectively
\begin{eqnarray}\label{solAzsmall2}
\xi(t)&\approx&\varsigma \sqrt{1+\varrho^2[\ln(t/t_0)]^2}\, ,\nonumber\\
S(t,x)&\approx&\arctan[\varrho\ln(t/t_0)] \pm k x \, ,
\end{eqnarray}
where $\varsigma$, $\varrho$ and $t_0$ are integration constants. This can be checked by inspection. In this case, the frequency is
\begin{equation}
K_0=\partial_0 S=\frac{\varsigma^2\varrho}{t\, \xi^2}\, ,
\end{equation}
and the wavevector $K_x=\partial_x S=k$. At very early times, the waves propagate along timelike curves
\begin{equation}
K^\mu K_\mu= -\frac{\varrho^2}{t^2\xi^4}+\frac{k^2}{2^{2/3}}<0\, ,
\end{equation}
as $t\rightarrow 0$.
The EM wave is dispersive as the dispersion relation depends on the wavelength $1/k\gg t$.

\subsubsection{$y$-polarized EM wave}
\label{subsectionypolarizedEMwave}

A similar approach can be done to study the $y$-polarized EM wave in the $x$-direction. In this case the  Eq.~\eqref{xAy} becomes
\begin{equation}\label{eqAyexplicita}
-\beta^{1/3}\Theta^{\frac{2}{3}}\partial_0\left(\beta^{1/3}\Theta^{-\frac{4}{3}}\partial_0 A_y\right)+\partial_x^2 A_y=0\, .
\end{equation}

 This equation can also be studied in the aforementioned two limits. For very late times we can use again the formalism developed in Sec.~\ref{modelteoricsimple}.
Using the same approximations than before, from  \eqref{solTeoejem6} we find that
\begin{equation}
\eta_y\approx -\left(\frac{\Lambda}{3}\right)^{3/2}\tau^3+ \left(\frac{\Lambda}{3}\right)^{3/2}\frac{k^2\tau^5}{5},
\end{equation}
and this wave has an amplitude and frequency given by
\begin{eqnarray}
\xi_y&\approx&\frac{\xi_{0z}}{\sqrt{k}}\left[1+ \left(\frac{\Lambda}{3}\right)^{3/2}\tau^3\right]\, ,\nonumber\\
K_0&\approx& \frac{k}{a}\left[1-\left(\frac{\Lambda}{3}\right)^{3/2}\tau^3\right]\, .
\end{eqnarray}
Thereby, this wave does not follow null geodesics, but
\begin{equation}\label{disp1lateb}
K_\mu K^\mu\approx\frac{2}{a^2}\left(\frac{\Lambda}{3}\right)^{3/2}k^2\tau^3>0\, .
\end{equation}
The EM plane wave has spacelike nature, approaching to a null geodesics behavior as $t\rightarrow \infty$.
It is very important to notice that the behavior \eqref{disp1lateb} of the $y$-polarized wave is different from the one of the $z$-polarized wave \eqref{disp1late}. This is a consequence of the anisotropy, and it produces birefringence.

For very early times, similar results can be found for the EM $y$-polarized wave along the $x$-direction. Eq.~\eqref{eqAyexplicita} becomes
\begin{equation}\label{vaccumAyearly}
-{2^{2/3}}{t}\partial_0\left(\frac{1}{t}\, \partial_0 A_y\right)+\partial_x^2A_y=0\, ,
\end{equation}
which has the exact solution
\begin{equation}\label{solHankel0small2222}
A_y(t,x)=A_{y0}\,t\,  H_1^{(1)}\left(2^{-1/3} k t\right) e^{\pm i k x}\, ,
\end{equation}
with an arbitrary constant $A_{y0}$, and the Hankel function of first order $H_1^{(1)}$. This wave presents a different behavior from the one described in \eqref{solHankel0small}, and therefore  at early times, the cosmological system is birefringent.
For very small times $k t\ll 1$, solution \eqref{solHankel0small2222} has the approximated form of a plane  wave
$A_y(t,x)=A_{y0} \xi(t) e^{i S(t,x)}$,
with the amplitude and phase given by
\begin{eqnarray}
\xi(t)&=&\beta+\frac{t^2}{\pi^2\beta}\left[1-\ln\left(t^2/t_0^2\right)\right]\approx \beta\, ,\nonumber\\
S(t,x)&=&-\arctan\left(\frac{2 \beta^2}{\varrho t^2}\right) \pm k x \, ,
\end{eqnarray}
where $\beta$ is an integration constant, and other constants previously defined.
As time approaches 
zero, the phase can be rewritten as $S = -\pi/2 +
\varrho t^2/(2\beta^2)$, and the frequency, when $t\rightarrow 0$, can be calculated as
\begin{equation}
K_0\approx \frac{\varrho t}{\beta^2}\, ,
\end{equation}
while the wavevector is $K_x=k$. Thereby, at very early times, the waves propagate along spacelike curves
\begin{equation}
K^\mu K_\mu\approx -\frac{\varrho^2t^2}{\beta^4}+\frac{k^2}{2^{2/3}}>0\, ,
\end{equation}
as $k\, t\rightarrow 0$.

The previous solutions show how anisotropy can modify the behavior of light, implying that EM waves move along timelike or spacelike curves as the Universe approaches to isotropization.
Also, for very early times, the large non--null geodesic behavior  is due to the extreme anisotropy that exist for very early times of the Universe. Both of these results show that there is a profound interplay between light propagation and gravity.

%%%%%%%%%%%%%%%%%%
\section{Redshift}

The non-null geodesic propagation of light seems to be a robust effect for anisotropic cosmologies. With the previous solutions for the four-wavevector $K_\mu$ we can also calculate the  cosmological redshifts measured by different co-moving observers who follow geodesics with four-velocity $U^\mu=(1,0,0,0)$.
 To perform this, we generalize the procedure presented in Ref.~\cite{carroll}.

Let us consider the following Killing tensor
\begin{eqnarray}\label{Killintensor}
{\cal K}_{\mu\nu} &=& h \left(g_{\mu\nu}+U_\mu U_\nu\right)+\Xi_a\, {\cal K}_\mu^{a}{\cal K}_\nu^{a}\nonumber\\
&&+\Xi_b\, {\cal K}_\mu^{b}{\cal K}_\nu^{b}+\Xi_c\, {\cal K}_\mu^{c}{\cal K}_\nu^{c}\, ,
\end{eqnarray}
that satisfies the equation $\nabla_\mu K_{\nu\sigma}+\nabla_\nu K_{\mu\sigma}+\nabla_\sigma K_{\nu\mu}=0$, and where
\begin{equation}
h=\frac{1}{3}\left(a^2+b^2+c^2\right)\, .
\end{equation}
depends on time only. Also, the Killing tensor \eqref{Killintensor} requires three time-dependent functions
\begin{eqnarray}
\Xi_a&=&\frac{2}{3}\int dt\,\, \frac{\left(b^2+c^2\right)\dot a-\left(b\dot b+c\dot c\right)a}{a^3}\, ,\nonumber\\
\Xi_b&=&\frac{2}{3}\int dt\,\, \frac{\left(a^2+c^2\right)\dot b-\left(a\dot a+c\dot c\right)b}{b^3}\, ,\nonumber\\
\Xi_c&=&\frac{2}{3}\int dt\,\, \frac{\left(a^2+b^2\right)\dot c-\left(a\dot a+b\dot b\right)c}{c^3}\, ,
\end{eqnarray}
that accompany the three Killing vectors for an anisotropic cosmology
\begin{eqnarray}\label{Killingvectors1}
{\cal K}_\mu^{a}&=&(0,a^2,0,0)\, ,\nonumber\\
{\cal K}_\mu^{b}&=&(0,0,b^2,0)\, ,\nonumber\\
{\cal K}_\mu^{c}&=&(0,0,0,c^2)\, .
\end{eqnarray}
Notice that in the isotropic limit $h=a^2$, $\Xi_1=0=\Xi_2=\Xi_3$, and the  Killing tensor \eqref{Killintensor} coincides with the one for the flat FRW cosmology \cite{carroll}.

As we showed in previous sections, we can find the solution for three different four-wavevectors, each one describing the wave propagation on each direction. These ones can be depicted as $K_{\mu}^{(1)}=(K_0^{(1)},k,0,0)$,
$K_{\mu}^{(2)}=(K_0^{(2)},0,k,0)$, and
$K_{\mu}^{(3)}=(K_0^{(3)},0,0,k)$, where $k$ is the constant associated to the spatial derivative of the phase. Here $K_0^{(j)}$ are three different frequencies, which differ due to the spacetime anisotropy ($j=1,2,3$ for propagation in the $x$, $y$ or $z$-directions respectively).  By Eq.~\eqref{SeparadaMaxGG3}, each  propagation defines $\chi_j=g^{\mu\nu}K_{\mu}^{(j)}K_{\nu}^{(j)}$.

The
Killing tensor \eqref{Killintensor} can be used to construct a scalar associated to the propagation of the wave. Thereby, three different scalars can be obtained as
\begin{equation}\label{constantredshift}
{\cal K}_j=g^{\mu\alpha}g^{\nu\beta}{\cal K}_{\mu\nu}K^{(j)}_{\alpha} K^{(j)}_\beta\, ,
\end{equation}
for the three different directions of propagation. These scalars have the property that they
are constants on the propagation along $K_\beta$,
\begin{eqnarray}
g^{\mu\nu}K_\mu^{(j)}\nabla_\nu{\cal K}_j=\partial_\mu \chi_j K_\nu^{(j)} {\cal K}^{\mu\nu}=\partial_0\chi_j K_\nu^{(j)} {\cal K}^{0\nu}=0\, .
\end{eqnarray}
This result holds because, from Eq.~\eqref{SeparadaMaxGG3} and Secs.~\ref{modelteoricsimple} and \ref{soluAnis},
 we realize that $\chi$ depends  only on time for this anisotropic model. Also, ${\cal K}^{0\nu}\equiv 0$, which can be proved directly from Eq.~\eqref{Killintensor}. Therefore, each scalar ${\cal K}^{(i)}$ is a constant that can be used to calculate the redshift  for EM plane waves \cite{carroll}. Moreover, as the constant scalar \eqref{constantredshift} depends on spacetime anisotropy (by the Killing vectors), then it will depend on the direction of propagation of the wave.

Now, let us assume a wave propagating on some $j$-direction.
By defining the  frequency of the wave propagating in the $j$-direction and measured by the co-moving observer as $\omega_j=-U^\mu K_\mu^{(j)}$, then any of the three constants \eqref{constantredshift} can be explicitly written as
\begin{equation}\label{constant222}
{\cal K}_j=h \chi_j+h \omega_j^2+ \Xi_j k^2\, .
\end{equation}
For any direction of propagation, Eq.~\eqref{constant222} allows to find the relation between an EM wave emitted with frequency $\omega_{j}(t_1)$ in a time $t_1$ and  the observed wave  frequency $\omega_{j}(t_2)$ measured in a time $t_2$
\begin{eqnarray}\label{redshiftsolutcion}
\left[\omega_{j}(t_2)\right]^2&=&\frac{h(t_1)}{h(t_2)}\left[\omega_{j}(t_1)\right]^2+\left(\frac{h(t_1)}{h(t_2)}\chi_j(t_1)-\chi_j(t_2)\right)\nonumber\\
&&\qquad+\frac{k^2}{h(t_2)} \left[\Xi_j(t_1)-\Xi_j(t_2)\right]\, .
\end{eqnarray}
This is the general redshift measured in an anisotropic spacetime for an EM plane wave beyond the geometrical optics limit. In general, there are three different redshifts, for each direction of propagation of the wave. Also, the redshift depends on the wave polarization through $\chi$ and $\Xi$. These results are in agreement with the coupling between polarization, direction of propagation and the gravitational field manifested in the previous sections.
We can use any solution of the previous section to calculate the redshift \eqref{redshiftsolutcion} for any direction of propagation.

Lastly, in the isotropic limit, we recover the usual redshift for FRW cosmologies \cite{carroll}
\begin{equation}
\omega(t_2)=\frac{a(t_1)}{a(t_2)}\omega(t_1)\, ,
\end{equation}
for light moving in null geodesics.

%%%%%%%%%%%%%%%%%%%%%%%%%%%%%
\section{Discussion}

The previous solutions show that the anisotropic spacetime acts as a fully birefringent medium. EM waves travelling on different directions have different propagations. It can also been shown that the direction of flux of energy of these waves behaves similar to waves in flat spacetime. The energy-momentum tensor of the EM field \cite{misner}
\begin{equation}
T^{\mu\nu}=F^{\mu\alpha}F_{\beta\alpha}g^{\beta\nu}-\frac{1}{4}g^{\mu\nu}F_{\alpha\beta}F^{\alpha\beta}\, ,
\end{equation}
is divergenceless, i.e., $\nabla_\nu T^{\mu\nu}=0$, from where  the flux of energy, described by the Poynting vector, can be calculated.

 First, for any direction of propagation of our solutions for EM waves, we can calculate the energy density of the wave
\begin{equation}
\left\langle T^{00}\right\rangle= (K^0)^2 \xi^2 +\frac{1}{2}g^{ii}\partial_0 \xi_i\partial_0 \xi_i+\frac{\chi}{2}\xi^2\, ,
\end{equation}
where  we have used the notation of Sec.~\ref{sec2sec}. Here, the energy-momentum tensor have been averaged over a wavelength. The last two terms represent corrections due to  the non-null geodesic behavior of light. Due to the explicit time dependence of the metric, this energy density is not a conserved quantity.

On the other hand, for any direction of propagation, we can define
the  Poynting vector $S^i$ as
\begin{equation}
S^i=\left\langle T^{0i}\right\rangle= K^0 \xi^2 K^i\, .
\end{equation}
 The  Poynting vector points along the same directions as the spatial wavevectors for any of the three different kinds of propagation.  Therefore, in our previous solutions, the flux of energy travels with the EM plane wave in the anisotropic cosmology, in an analogue fashion to what occurs on flat spacetime.

%%%%%%%%%%%%%%%%%%%%%%%%%%%%%
Therefore, we have shown that the EM plane waves discussed along this work represent physical solutions to Maxwell equations, that  can be tested using the cosmological redshift \eqref{redshiftsolutcion}.
These results may be useful for giving a new perspective on some outstanding problems in early--Universe and inflationary cosmology \cite{guth,linde,albre}. There are several approaches aiming to explain open inflationary questions using variable speed of light (VSL) theories  \cite{kragh,moffat,Magu1,Magu2,Magu3,petit1,moffat2}. To the best of our knowledge, no convincing mechanisms have been proposed to explain the VSL behavior. However, in an anisotropic cosmology, we have shown that naturally there exist light wave solutions  with spacelike propagation (Sec.~\ref{subsectionypolarizedEMwave}). Thus,
the  velocity of those EM waves at early times can be very high in such  way that there is no need for extra (scalar) fields to solve the so called ``horizon'' problem \cite{afshor}. For latter times, in addition to the already known isotropization of the background, the geodesic behavior and  speed of light can reach the current values, thus solving one of the outstanding inflation problems within the realm of unmodified General Relativity.
 This can be related with  symmetry properties of the equations of Sec.~\ref{Propagationoflightinanisotropicsec}, which are discussed in the Appendix.

We believe that this work can bring some new light on these old problems.

\begin{acknowledgments}

F.A.A. thanks the CONICyT-Chile for Funding N$^o$ 79130002. The authors wish to express their gratitude to the anonymous referee for very useful comments and suggestions.

\end{acknowledgments}

\appendix
\section{Anisotropic symmetries of Maxwell equations}
\label{dddax}

The transverse components of Maxwell equations \eqref{yAxG} may be written in the generic fashion (for  propagation in any direction) as
\begin{equation}
F G A_\beta=\partial_\alpha^2A_\beta\, ,\quad G F A_\gamma=\partial_\alpha^2A_\gamma\, ,\label{SuSy22}
\end{equation}
with the operators $F \equiv f(t)\partial_0$ and $G \equiv g(t)\partial_0$.
Eqs.~\eqref{SuSy22} exhibit a new kind of symmetry which is reminiscent of (but not equivalent to) supersymmetry in the sense of one dimensional quantum mechanics as realized by Darboux \cite{darboux} and Witten \cite{witten}. In fact, multiplying Eqs.~\eqref{SuSy22} by $G$ and $F$, respectively, we get
\begin{equation}
G F G A_\beta=G \partial_\alpha^2A_\beta\, , \quad F G F A_\gamma=F \partial_\alpha^2A_\gamma\, .\label{SuSy222}
\end{equation}
It is straightforward to realize that $G A_\beta$ and $F A_\gamma$ satisfy the corresponding Eqs.~\eqref{SuSy22} because $F$ and $G$ commute with $\partial_\alpha^2$ .
Let us just consider Eqs.~\eqref{xAy} and \eqref{xAz} for propagation of EM wave in the $x$-direction. The relations between the different type of polarization states are
\begin{equation}\label{symme}
A_z=\frac{a c}{b}\, {\partial_0 A_y}\, ,\qquad A_y=\frac{a b}{c}\, \partial_0 A_z\, .
\end{equation}
Therefore, solving the equation for one polarization state allow us to find the behavior of the other one.
Similar symmetry relations apply to the other directions of propagation defined by  Eqs.~\eqref{yAxG}.


\begin{thebibliography}{}

\bibitem{pebles} P. J. E. Peebles, Nuclear Physics B (Proc. Suppl.) {\bf 138}, 5 (2005).
\bibitem{inflation} A. R. Liddle and D. H. Lyth, {\it Cosmological Inflation and Large-Scale Structure} (Cambrigde University Press, 200).
\bibitem{misner} C. W. Misner, K. S. Thorne and J. A. Wheeler, {\it Gravitation} (W. H. Freeman and Co. San Francisco, 1973).
\bibitem{Whitt} E. T. Whittaker, Math. Proc. Cambr. Phil. Soc. {\bf 24}, 32 (1928).
\bibitem{mash1987} B. Masshoon, Phys. Lett. A {\bf 122}, 299 (1986).
\bibitem{dwb} B. S. DeWitt and R. W. Brehme, Ann. Phys. (N.Y.) {\bf 9}, 220 (1960)
\bibitem{Faraoni} V. Faraoni and S. Sonego, Phys. Lett. A {\bf 170}, 413 (1992).
\bibitem{asenjohojmanLight} F. A. Asenjo and S. A. Hojman,  {\it Do electromagnetic waves always propagate along null geodesics?}, arXiv:1608.06572 (2016).
\bibitem{ohkuwa} Y. Ohkuwa, Prog. Theor. Phys. {\bf 65}, 1058 (1981).
\bibitem{velo} G. Velo and D. Zwanziger, Phys. Rev. {\bf 186}, 1337 (1969).
\bibitem{Drummond} I. T. Drummond and S. J. Hathrell, Phys. Rev. D {\bf 22}, 343 (1980).
\bibitem{Klein} J. J. Klein and B. P. Nigam, Phys. Rev. {\bf 135}, B1279 (1964).
 \bibitem{danielshore} R. D. Daniels and G .M. Shore, Nuc. Phys. B {\bf 425} 634 (1994).
\bibitem{danielshore2} R. D. Daniels and G .M. Shore,  Phys.Lett.  B {\bf 367} 75 (1996).
\bibitem{danielshore3} G .M. Shore,  Nuc. Phys.  B {\bf 460} 379 (1996).
\bibitem{danielshore4} G .M. Shore,  Nuc. Phys.  B {\bf 633} 271 (2002).
\bibitem{Khriplovich} I. B. Khriplovich, Phys. Lett. B {\bf 346}, 251 (1995).
\bibitem{khri} I. B. Khriplovich, Phys. Lett. B {\bf 346}, 251 (1995).
\bibitem{hollow} T. J. Hollowood and G. M. Shore, J. High Energy Phys. {\bf 2008}, 091 (2008).
\bibitem{heylH} J. S. Heyl and L. Hernquist, Jour. Phys. A: Math. Gen. {\bf 30}, 6485 (1997)
\bibitem{agullo} I. Agullo, A. del Rio and J. Navarro-Salas, Phys. Rev. Lett. {\bf 118}, 111301 (2017).
\bibitem{mat} M. Mathisson, Acta Physica Polonica {\bf 6}, 163 (1937); {\bf 6}, 218 (1937).
\bibitem{pap1} A. Papapetrou, Proc. Roy. Soc (London) {\bf 209A}, 248 (1951).
\bibitem{hojman1} S. A. Hojman, {\it Electromagnetic and Gravitational Interactions of a Spherical Relativistic Top},
       Ph.D. thesis,Princeton University, 1975 (unpublished).
\bibitem{gane1} S. A. Hojman and F. A. Asenjo, Class. Quantum Grav. {\bf 30}, 025008 (2013).
\bibitem{gane2} N. Zalaquett, S. A. Hojman and F. A. Asenjo, Class. Quantum Grav. {\bf 31}, 085011 (2014).
\bibitem{armaza} C. Armaza, S. A. Hojman, B. Koch and N. Zalaquett, Class. Quantum Grav. {\bf 33}, 145011 (2016).
\bibitem{evan} E. A. Evangelidis, Astroph. Space Sci. {\bf 60}, 213 (1979).
\bibitem{mashhoonPRD} B. Mashhoon, Phys. Rev. D {\bf 7}, 2807 (1973).
\bibitem{ruizdodin} D. E. Ruiz and I. Y. Dodin, Phys. Plasmas {\bf 24}, 055704 (2017).
\bibitem{mackay} T. G. Mackay, A. Lakhtakia and S. Setiawan, New J. Phys. {\bf 7}, 75 (2005).
\bibitem{felice} F. de Felice, Gen. Rel. Grav. {\bf 2}, 374 (1971).
\bibitem{mashhoon} B. Mashhoon, Phys. Rev. D {\bf 8}, 4297 (1973).
\bibitem{mashhoon2} B. Mashhoon, Phys. Rev. D {\bf 10}, 2679 (1975).
\bibitem{plebanski} J. Plebanksi, Phys. Rev. {\bf 118}, 1396 (1960).
\bibitem{brans} C. H. Brans, ApJ {\bf 197}, 1 (1975).
\bibitem{sagnotti} A. Sagnotti and B. Zwiebach, Phys. Rev. D {\bf 92}, 305 (1981)
\bibitem{Moskaliuk} S. S. Moskaliuk, Phys. Rev. D{\bf 68}, 084023 (2003).
\bibitem{kasner21} E. Kasner,  Am. J. Math. {\bf 43}, 217 (1921).
\bibitem{ryan} M. P. Ryan, Jr. and L. C. Shepley, {\it  Homogeneous Relativistic Cosmologies} (Princeton University Press, Princeton, 1975).
\bibitem{goshsen} T. Ghosh and A. K. Sen, ApJ. {\bf 833}, 82 (2016).
\bibitem{teu} S. A. Teukolsky, Phys. Rev. Lett. {\bf 29}, 1114 (1972).
\bibitem{chandra}S. Chandrasekhar, Proc. R. Soc. Lond. A {\bf 349}, 1 (1976).
\bibitem{cohen} J. M. Cohen and L. S. Kegeles, Phys. Rev. D {\bf 10}, 1070 (1974).
\bibitem{tsagas} C. G. Tsagas, Class. Quantum Grav. {\bf 22} 393 (2005).
\bibitem{born} M. Born and E. Wolf, {\it Principles of Optics, 6th ed.} (Pergamon Press, 1980).
\bibitem{tolman} R. C. Tolman, {\it Relativity, Thermodynamics and Cosmology} (Dover Publications Inc. New York, 1987).
\bibitem{noonan} T. W. Noonan, Class. Quantum Grav. {\bf 12}, 1087  (1995).
\bibitem{Gron} {{\O}}. Gr{{\o}}n and S. Hervik, {\it Einstein's General Theory of Relativity: with modern applications in cosmology} (Springer, 2007).
\bibitem{carroll} S. Carroll, {\it Spacetime and Geometry, An introduction to General Relativity} (Addison Wesley, San Francisco, 2004).

\bibitem{guth} A. H. Guth, Phys. Rev. D {\bf 23}, 347 (1981).
\bibitem{linde} A. D. Linde, Phys. Lett. {\bf 108B}, 389 (1982).
\bibitem{albre} A. Albrecht and P. J. Steinhardt, Phys. Rev. Lett. {\bf 48}, 1220 (1982).	
\bibitem{kragh} H. S. Kragh, Studies in History and Philosophy of
Modern Physics {\bf 37}, 726 (2006).
\bibitem{moffat} J. W. Moffat, Int. J. Mod. Phys. D {\bf 2}, 351 (1993).
\bibitem{petit1} J.-P. Petit, Mod. Phys. Lett. A {\bf 3}, 1527 (1988).
\bibitem{Magu1} A. Albrecht and J. Magueijo, Phys. Rev. D {\bf 59}, 043516 (1999).
\bibitem{Magu2} J. Magueijo, Phys. Rev. D {\bf 62}, 103521 (2000).
\bibitem{Magu3} J. Magueijo, Rep. Pro. Phys. {\bf 66}, 2025 (2003).
\bibitem{moffat2} J. W. Moffat, Eur. Phys. J. C {\bf 76}, 130 (2016).
\bibitem{afshor} N. Afshordi and J. Magueijo, Phys. Rev. D {\bf 94}, 101301(R) (2016).
\bibitem{darboux} G. Darboux, C.R. Acad. Sci. {\bf 94}, 1456 (1882).
\bibitem{witten} E. Witten, Nucl Phys. B {\bf 188}, 513 (1981).



\end{thebibliography}
\end{document}